\documentstyle[12pt]{article}
\title{\bf One-Loop Free Energy\\
of the Four-Dimensional Compact QED\\
in the Confining Phase}
\author{D.V.ANTONOV \thanks{E-mail addresses: 
antonov@pha2.physik.hu-berlin.de and antonov@vxitep.itep.ru}{\,}
\thanks{Supported by Graduiertenkolleg {\it Elementarteilchenphysik}, 
Russian Fundamental Research Foundation, Grant No.96-02-19184, DFG-RFFI, 
Grant 436 RUS 113/309/0, and by the INTAS, Grant No.94-2851.}
\\
{\it Institute of Theoretical and Experimental Physics,}\\
{\it B.Cheremushkinskaya 25, 117 218, Moscow, Russia}\\
{\it and}\\
{\it Institut f\"ur Physik, Humboldt-Universit\"at zu Berlin,}\\
{\it Invalidenstrasse 110, D-10115, Berlin, Germany}} 
\date{}
\begin{document}
\maketitle
\vspace{1cm}
\centerline{\bf {Abstract}}

\vspace{3mm}
The one-loop free energy of the four-dimensional compact QED, 
which is known to be equivalent to the vector Sine-Gordon model, 
is calculated in the strong coupling regime. In the case, 
when the norm of the strength tensor 
of the saddle-point 
value of the corresponding Sine-Gordon model is much larger than the 
typical inverse area of a loop in the gas of the monopole rings, 
the obtained free energy decays exponentially 
vs. this norm. In the opposite case, when 
the dominant 
configuration of the Sine-Gordon model is identically zero, 
the resulting free energy decays with the growth 
of loops as an exponent of the inverse square of their typical area.     

\newpage

It is known [1,2], that the so-called strong coupling regime of the 4D 
compact QED, when its coupling constant $g$ is larger than some critical 
value, corresponds to the confining phase of this theory. 
In the present Letter, we shall calculate the one-loop free energy of 
the 4D compact QED in such a case, i.e. in the case when $g\gg 1$. 
The expression for the partition 
function of this model has been found in Ref. [1] and has the following form

$$Z\left(g^2\right)_{\rm 1-loop}\equiv {\rm e}^{-{\cal F}}=
Z\left(g^2\right)_{\rm 0-loop}\int 
{\cal D}\chi_\mu\exp\Biggl\{\int d^4x\Biggl[\frac{g^2}{16\pi^2}
\chi_\mu\Box\chi_\mu+$$

$$+\frac{2}{a^4}\int {\cal D}x_\mu(t)
\exp\left(-\Sigma_{\alpha\sigma}\Sigma_{\beta\sigma}G_{\alpha\beta}
\right)\cos\left(\Sigma_{\mu\nu}\partial_{[\mu}\chi_{\nu]}\right)
\Biggr]\Biggr\}. \eqno (1)$$
Here $\frac{1}{a}$ is an UV momentum cutoff, $G_{\alpha\beta}\equiv
\frac{S_{\alpha\beta}^{(0)}}{g^2a^4}$, where $S^{(0)}$ is a dimensionless 
constant symmetric tensor, and $\Sigma_{\alpha\beta}\equiv\oint
\limits_C^{} x_\alpha 
dx_\beta$ 
is the so-called tensor area [3] associated with the contour $C:~
x_\alpha(t),~ 0\le t\le T,~ x_\alpha(0)=x_\alpha(T)$. 
Notice that $\Sigma_{\alpha
\sigma}\Sigma_{\beta\sigma}G_{\alpha\beta}$ is just the 
self-action of a single monopole loop, whereas Eq. (1) is the partition 
function of a gas of such  
loops, where in the confining phase under consideration the 
large loops are dominant, i.e. this gas is not dilute.  

Without loss of generality, the symmetric matrix $G_{\alpha\beta}$ could 
be chosen to be a diagonal one. Then, since 

$$\Sigma_{\alpha\sigma}\Sigma_{\beta\sigma}G_{\alpha\beta}=\oint\limits_C^{} 
dx_\sigma\oint\limits_C^{} 
dy_\sigma x_\alpha y_\beta G_{\alpha\beta}=-\frac12 
\sum\limits_{\alpha=1}^{4}G_{\alpha\alpha}\oint\limits_C^{} 
dx_\sigma\oint\limits_C^{} dy_\sigma 
\left(x_\alpha-y_\alpha\right)^2,$$
one can see that $\left(G_{\alpha\alpha}\right)^{-\frac14}$ is a typical 
size of a monopole loop in the $\alpha$-direction (of course, all the 
four components of the matrix $G_{\alpha\beta}$ are positive by definition). 
In what follows, we shall consider only the loops whose sizes along at 
least three axes, $x,y$, and $z$, are equal to each other, so that the 
matrix $G_{\alpha\beta}$ has the form $G_{\alpha\beta}={\rm diag}\left( 
\lambda_1,\lambda_1,\lambda_1,\lambda_2\right)$. Then one can linearize 
the term bilocal in the tensor area in Eq. (1) by introducing the integration 
over an auxiliary antisymmetric matrix\footnote{Such a linearization 
for the case when $G_{\alpha\beta}\sim\delta_{\alpha\beta}$ has been 
first proposed in [4].}  as follows

$${\rm e}^{-\Sigma_{\alpha\sigma}\Sigma_{\beta\sigma}G_{\alpha\beta}}=
\frac{1}{2\sqrt{2}\pi^3\lambda_1^{\frac32}\left(\lambda_1+\lambda_2
\right)^{\frac32}}\int dB 
{\rm e}^{B_{\alpha\beta}A_{\beta\gamma}
B_{\gamma\alpha}-iB_{\alpha\beta}\Sigma_{\alpha\beta}}, \eqno (2)$$
where 

$$\int dB\equiv\int\limits_{-\infty}^{+\infty} dB_{12}dB_{13}
dB_{14}dB_{23}dB_{24}dB_{34},$$
and

$$A_{\alpha\beta}=\frac{1}{4\lambda_1}{\rm diag}\left(1,1,1,
\frac{3\lambda_1-\lambda_2}{\lambda_1+\lambda_2}\right).$$
Notice, that since both $\lambda_1$ and $\lambda_2$ are positive, there 
could not be any singularities neither in the matrix $A_{\alpha\beta}$, nor 
in the denominator on the R.H.S. of Eq. (2). 

In the regime we are interested with, when $g\gg 1$, the term 

$$\exp\left(\frac{g^2}{16\pi^2}\int d^4x\chi_\mu\Box\chi_\mu\right)$$
in the functional integral on the R.H.S. of Eq. (1) is dominant. 
Therefore, up to an inessential constant, the one-loop free energy  
following from Eq. (1) will be equal to the value of the expression,   
standing in the 
second line of this equation, which it acquires at the saddle-point 
of the integral

$$\int {\cal D}\chi_\mu\exp\left(\frac{g^2}{16\pi^2}\int d^4x\chi_\mu
\Box\chi_\mu\right).$$
This saddle-point value obviously reads $\chi_\mu=\frac12 x_\nu F_{\nu\mu}$, 
where $F_{\mu\nu}$ is a constant antisymmetric matrix, which in particular 
could be identically zero. Hence,

$${\cal F}=-\frac{1}{2\sqrt{2}\pi^3a^4\lambda_1^{\frac32}\left(\lambda_1+
\lambda_2\right)^{\frac32}}\int dB {\rm e}^{B_{\alpha\beta}A_{\beta\gamma}
B_{\gamma\alpha}}\int {\cal D}x_\mu(t){\rm e}^{-i B_{\alpha\beta}
\Sigma_{\alpha\beta}}\left({\rm e}^{i F_{\alpha\beta}\Sigma_{\alpha\beta}}+
{\rm e}^{-i F_{\alpha\beta}\Sigma_{\alpha\beta}}\right). \eqno (3)$$
In order to calculate the integral over contours standing on the R.H.S. of 
Eq. (3), let us regularize it by introducing a small kinetic term 
${\rm e}^{-\frac{1}{4g^2}\int\limits_0^T dt\dot x^2}$, after which this 
integral becomes nothing else, but the one-loop unsubtracted 
Euler-Heisenberg-Schiwinger Lagrangian in scalar QED [5], 
and Eq. (3) yields [6,7] 

$${\cal F}=-\frac{g^4}{8\sqrt{2}\pi^5a^4\lambda_1^{\frac32}
\left(\lambda_1+\lambda_2\right)^{\frac32}}\int dB {\rm e}^{B_{\alpha\beta}
A_{\beta\gamma}B_{\gamma\alpha}}\Biggl[\frac{a_1b_1}{\sin\left(2g^2a_1T
\right)\sinh\left(2g^2b_1T\right)}+$$

$$+\frac{a_2b_2}{\sin\left(2g^2a_2T\right)
\sinh\left(2g^2b_2T\right)}\Biggr], \eqno (4)$$
where 

$$a_i^2=\frac12\left[\vec H_i^2-\vec E_i^2+\sqrt{\left(\vec H_i^2-\vec E_i^2
\right)^2+4\left(\vec E_i\vec H_i\right)^2}\right],$$

$$b_i^2=\frac12\left[\vec E_i^2-\vec H_i^2+\sqrt{\left(\vec H_i^2-\vec E_i^2
\right)^2+4\left(\vec E_i\vec H_i\right)^2}\right],$$
$i=1,2$. Here $\vec E_1, \vec B_1$ and $\vec E_2, \vec B_2$ stand for the 
``electric'' 
and 
``magnetic fields'' corresponding to the ``field strength tensors'' 
$F_{\alpha\beta}-B_{\alpha\beta}$ and $F_{\alpha\beta}+B_{\alpha\beta}$ 
respectively. Therefore we have replaced the integral over contours 
standing in the initial expression (1) by the averaging over an auxiliary 
matrix $B_{\alpha\beta}$ with a certain Gaussian weight, which is obviously 
much more simple.

Let us now proceed with performing this average in the simplest case when 
the loops are flat, i.e. only two components of the tensor area (which 
for concreteness will be chosen to be the components $\Sigma_{23}$ and 
$\Sigma_{32}$ respectively)  
are nonzero. In this case, Eq. (4) yields [7]

$${\cal F}=-\frac{g^2}{8\pi^2\sqrt{2\pi\lambda_1}Ta^4}\int
\limits_{-\infty}^{+\infty}dx {\rm e}^{-\frac{x^2}{2\lambda_1}}
\left[\frac{x+F}{\sinh\left(2g^2T\left(x+F\right)\right)}+
\frac{x-F}{\sinh\left(2g^2T\left(x-F\right)\right)}\right], \eqno (5)$$
where $F\equiv F_{23}$ is positive by definition. Since $g\gg 1$, in what 
follows we shall always assume that $g^2T\sqrt{\lambda_1}\gg 1$. 

The integral on the R.H.S. of Eq. (5) can be most easily carried out 
in the two following cases. First of them is the case of the 
strong field $\chi_\mu$, when $F\gg\sqrt{\lambda_1}$. In this case, one finds

$${\cal F}=\frac{g^2\left(2g^2T\lambda_1-F\right)
{\rm e}^{-2g^2FT}}{2\pi^2Ta^4},$$
which means the exponential falloff of the free energy 
vs. the norm $F$ of the 
strength tensor $F_{\alpha\beta}$. On the other hand, we get the restriction 
for the upper limit of $F$, which follows from the requirement of the 
positiveness of the obtained free energy, $F<2g^2T\lambda_1$. This means 
that the saddle-point configurations of the vector Sine-Gordon theory, 
we are studying, with too large strengthes are unstable. 

The other case is the degenerate one, when $F=0$. In this case, 

$${\cal F}=\frac{g^4\lambda_1{\rm e}^{2g^4T^2\lambda_1}}{\pi^2a^4},$$
which means that without the scale entering the system via the field 
$\chi_\mu$, 
the free energy decays with the growth of loops as an exponent of the 
inverse square of their typical area.

\vspace{3mm}
{\large\bf References}

\vspace{3mm}
\noindent
$[1]$~P.Orland, {\it Nucl.Phys.} {\bf B205[FS5]}, 107 (1982).\\
$[2]$~A.M.Polyakov, {\it Gauge Fields and Strings} (Harwood Academic 
Publishers, 1987).\\ 
$[3]$~A.M.Polyakov, {\it Phys.Lett.} {\bf B82}, 247 (1979), {\it 
Nucl.Phys.} {\bf B164}, 171 (1980); M.A.Shifman, 
{\it Nucl.Phys.} {\bf B173}, 13 (1980); A.A.Migdal, 
{\it Int.J.Mod.Phys.} {\bf A9}, 1197 (1994).\\
$[4]$~K.L.Zarembo, privat communication.\\
$[5]$~C.Itzykson and J.Zuber, {\it Quantum Field Theory} (McGraw-Hill, 
1980).\\
$[6]$~M.G.Schmidt and C.Schubert, {\it Phys.Lett.} {\bf B318}, 438 (1993).\\
$[7]$~M.Reuter, M.G.Schmidt, and C.Schubert, {\it Ann.Phys.} {\bf 259}, 
313 (1997).

\end{document}